\documentclass[pra,reprint,showpacs,superscriptaddress,showkeys,showpacs,times]{revtex4-1}
\usepackage{amsmath,amssymb,amsthm,times,graphics,graphicx,bm,subfigure}
\usepackage[colorlinks,linkcolor=blue]{hyperref}

\newcommand{\be}{\begin{equation}}
\newcommand{\ee}{\end{equation}}
\newcommand{\ben}{\begin{eqnarray}}
\newcommand{\een}{\end{eqnarray}}
\newcommand{\bes}{\begin{subequations}}
\newcommand{\ees}{\end{subequations}}
\newcommand{\bF}{\begin{figure}}
\newcommand{\eF}{\end{figure}}

\newcommand{\arxiv}[2][arxiv:]{\href{http://arxiv.org/abs/#1#2}{#1#2}}

\def\ket#1{ | #1 \rangle}
\def\bra#1{{\langle #1 |  }}

\begin{document}
\title{A compact entanglement distillery}

\author{Animesh Datta}
\email{animesh.datta@physics.ox.ac.uk}
\affiliation{Clarendon Laboratory, Department of Physics, University of Oxford, OX1 3PU, United Kingdom}

\author{Lijian Zhang}
\affiliation{Clarendon Laboratory, Department of Physics, University of Oxford, OX1 3PU, United Kingdom}

\author{Joshua Nunn}
\affiliation{Clarendon Laboratory, Department of Physics, University of Oxford, OX1 3PU, United Kingdom}

\author{Nathan K. Langford}
\affiliation{Clarendon Laboratory, Department of Physics, University of Oxford, OX1 3PU, United Kingdom}

\author{Alvaro Feito}
\affiliation{ Vestas Technology Research \& Development, Venture Quays, East Cowes, PO32 6EZ, United Kingdom}

\author{Martin B. Plenio}
\affiliation{Institut f\"{u}r Theoretische Physik, Albert-Einstein-Allee 11, Universit\"{a}t Ulm, D-89069 Ulm, Germany}
\affiliation{QOLS, Blackett Laboratory, Imperial College London, Prince Consort Rd., SW7 2BW, United Kingdom}

\author{Ian A. Walmsley}
\affiliation{Clarendon Laboratory, Department of Physics, University of Oxford, OX1 3PU, United Kingdom}

\date{\today}


\begin{abstract} 
Large-scale quantum-correlated networks could transform technologies ranging from communications and cryptography to computation, metrology, and simulation of novel materials. Critical to achieving such quantum enhancements is distributing high-quality entanglement between distant nodes. This is made possible in the unavoidable presence of decoherence by entanglement distillation. However, current versions of this protocol are prohibitively costly in terms of resources. We introduce a new scheme for continuous-variable entanglement distillation that requires only linear temporal and constant physical or spatial resources, both of which are exponential improvements over existing protocols.  Our scheme uses a fixed module –- an \emph{entanglement distillery} –- comprising only four quantum memories of at most 50\% storage efficiency and allowing a feasible experimental implementation. Tangible quantum advantages are obtained by using non-ideal quantum memories outside their conventional role of storage. By creating, storing and processing information in the same physical space, the scheme establishes a potentially valuable technique for designing stable, scalable protocols across different quantum technologies.
\end{abstract}


\maketitle

Using the same physical space for storing and processing information is an attractive paradigm for material and information science~\cite{dietl10}. It will make devices smaller, and increase robustness and speed by dispensing with the necessity of moving information around. Such a development is particularly promising for quantum information science, where stored quantum states awaiting processing are extremely vulnerable to noise~\cite{Chuang_LSZ95,divincenzo95}, and shuttling information-bearing quantum states around a quantum device presents formidable challenges. Although enormous progress is being made in manipulating quantum information in a variety of systems and scenarios~\cite{Leanhardt_CKSGKP02,hensinger_OSHYADM06,beugnon,home_HJALW09}, theoretical proposals are necessary that minimize the demands on such manipulation. This is vital if the resources and advantages offered by quantum physics, in applications ranging from communications to metrology~\cite{Duan_LCZ01,Giovanetti_LM11}, are to be harnessed. Preparing distributed quantum states with high quality quantum entanglement lies at the heart of this endeavour. Unfortunately, noise and decoherence inevitably deteriorate the quality of entanglement in states that are shared between distant parties~\cite{NC00}. The situation can be remedied by entanglement distillation, in which multiple copies of a less entangled state are transformed by local operations and classical communications (LOCC) into fewer copies of a more entangled state~\cite{dist}.

\begin{figure}[h]
\begin{center}
\includegraphics[width=\columnwidth]{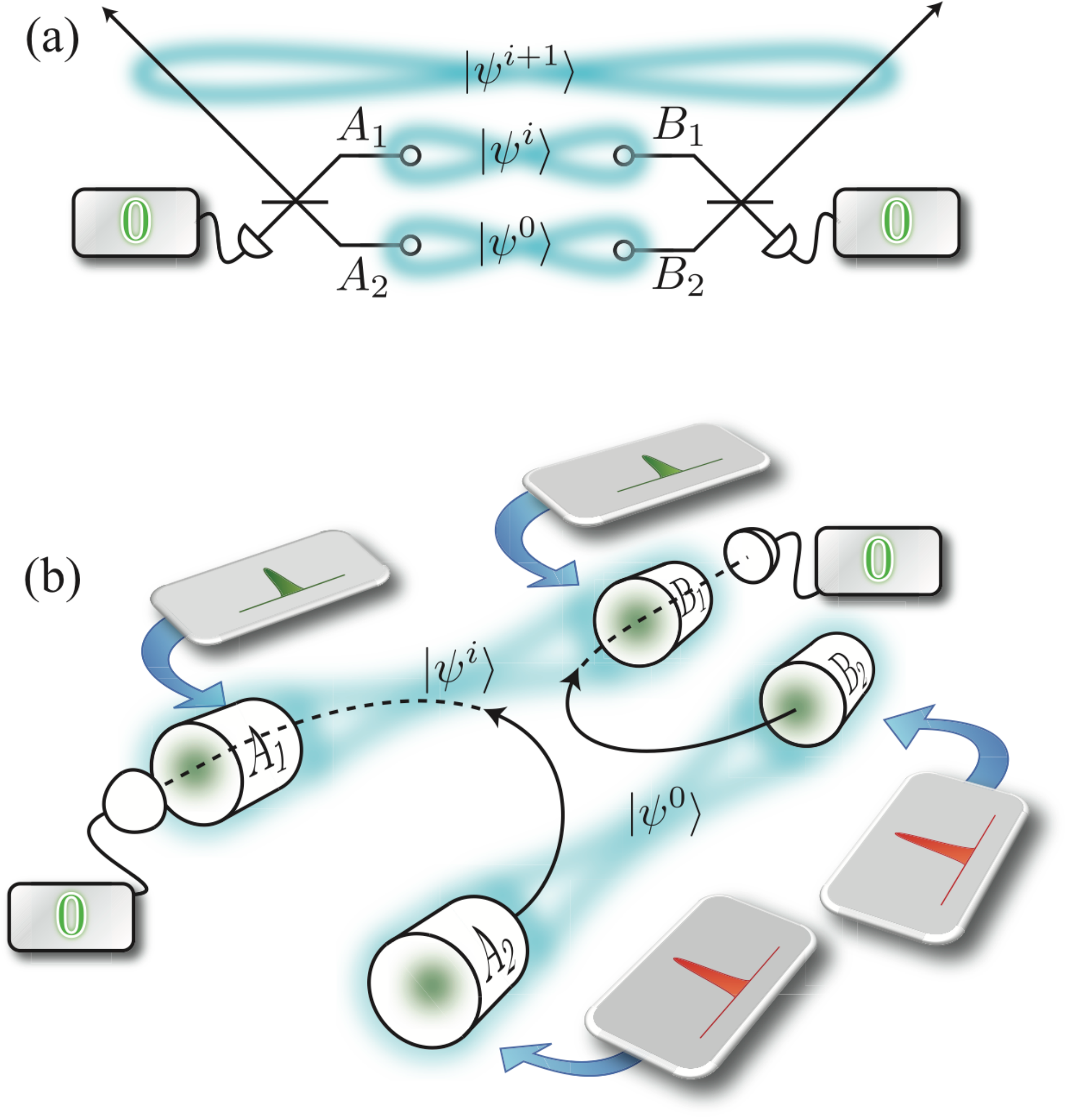}
\caption{Mashing. (a) A linear-optics schematic of the iterative mashing protocol, in which an entangled resource state $\ket{\psi^0}$, distributed between two parties Alice and Bob, is interfered with the shared entangled state $\ket{\psi^i}$ on 50:50 beamsplitters. Detection of vacuum by Alice and Bob heralds the success of the protocol, which produces a more entangled state $\ket{\psi^{i+1}}$. (b) Implementation of mashing using four quantum memories. The resource state is generated between memories $A_2$ and $B_2$, while the state $\ket{\psi^{i}}$ is shared between memories $A_1$ and $B_1$. The grey panels show the control pulses required to drive the memory interactions: full retrieval (red) and 50:50 beamsplitter (green). Mashing is achieved by retrieving the resource state from $A_2$, $B_2$, and sending it through memories $A_1$, $B_1$ while driving a 50:50 beamsplitter interaction, as described in the main text. Vacuum detections herald the successful production of $\ket{\psi^{i+1}}$ between $A_1$ and $B_1$ (not shown).}
\label{fig:distillation_mashing}
\end{center}
\end{figure}

\begin{figure}[h]
\begin{center}
\includegraphics[width=\columnwidth]{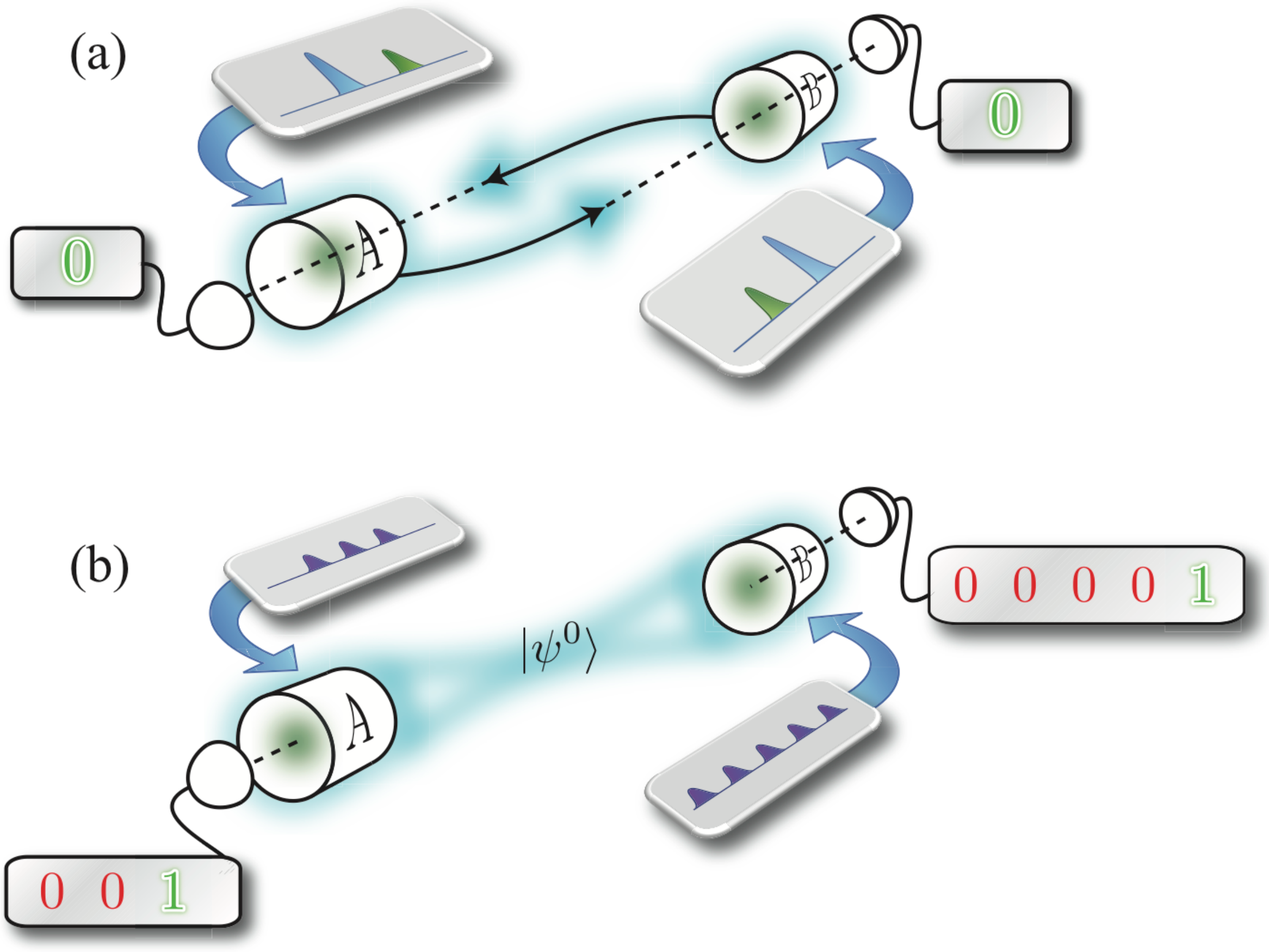}
\caption{Malting using quantum memories. The grey panels show the pulses required to drive the memory interactions: squeezing (blue), 50:50 beamsplitter (green), weak beamsplitter (purple). (a) A squeezing interaction is driven in both memories $A$ and $B$, which each emit photons entangled with spin-wave excitations. The propagating photons from $A$ are directed into $B$ and vice versa, while a 50:50 beamsplitter interaction is driven, causing the photons and spin-wave excitations to interfere. Vacuum detections herald the generation of a two-mode squeezed state shared by $A$ and $B$. (b) De-Gaussification is achieved by photon subtraction at $A$ and $B$. Because the states are stored in the memories, a repeat-until-success strategy can be employed. Weak control pulses drive a series of beamsplitter interactions with small effective reflectivities. When a retrieved photon is detected at both $A$ and $B$, the malting process is over and the non-Gaussian entangled resources state $\ket{\psi^0}$ has been successfully generated.}
\label{fig:distillation_malting}
\end{center}
\end{figure}

We introduce the notion of a quantum entanglement distillery. For continuous variable (CV) entanglement, it provides improved use of fixed resources to achieve the same levels of improvement in entanglement as earlier schemes. In fact, our distillery has doubly-exponential temporal and exponential spatial advantage over existing distillation schemes~\cite{Browne_ESP03,Eisert_BSP04}. It also surpasses crucial limitations of finite-dimensional entanglement pumping schemes~\cite{Dur_B07,Pan_SBZ01}. In particular, failed local operations merely reduce the entanglement of the states involved as opposed to a finite-dimensional instance where failed attempts lead to completely unentangled states. Ours is a repeat-until-success scheme using a linear number of initially poorly entangled states to obtain a final state with higher entanglement. The distillery is a fixed module, consisting of only four quantum memories where the final amount of entanglement is governed by the initial states and the number of iterations. Memories~\cite{simonetal10} allow us to store results from previous iterations while the subsequent ones succeed, providing an exponential advantage in time. The additional exponential advantage in space and time is provided by our entanglement distillation protocol which we describe later.

The quantum memories not only store quantum information, but also process it concurrently in the same physical space. They allow us to repeatedly perform probabilistic operations on the same copy of the quantum state, further saving time and enhancing resilience against decoherence. This is vital as all local schemes for distilling entanglement must be probabilistic, since entanglement cannot, on average, increase under LOCC. Furthermore, distillation of CV entanglement is not possible if all states and operations involved are Gaussian~\cite{Eisert_SP02,Fiurasek02,Giedke_C02}. This can be circumvented by performing probabilistic non-Gaussian operations on initially Gaussian states~\cite{Browne_ESP03,Eisert_BSP04,Eisert_SP02,Fiurasek02,Giedke_C02,sasakidist}. A major advantage over finite-dimensional schemes is that failed local operations do not require starting the whole process anew. Finally, our scheme is also event-ready, in that the protocol's success is reported by fixed detector outcomes.

There are two key features the quantum memories must possess to be suitable for our distillery. Firstly, their time-bandwidth product, which determines the number of iterations that can be executed within its coherence lifetime, must be sufficiently large. Secondly, they need a transparent failure mode, i.e. they transmit any unstored excitation, allowing them to be used as a beamsplitter, and enabling \textit{in-situ} generation of the initial two-mode squeezed state.  Most importantly, however, at no point do we require a perfect memory.

We begin by describing our protocol, consisting of two major steps we call malting and mashing. We first describe mashing, which provides iterative improvement of a weak entanglement resource, and then the design of our quantum entanglement distillery that implements our scheme. Then we describe malting, the generation of a weak entangled resource, and analyze its success probability. Finally, we show how realistic memories allow us sufficiently many iterations to get very close to the limiting state.

\section*{Mashing}

The mashing step of entanglement distillation begins with a non-Gaussian resource state $\ket{\psi^0}$. This state is produced in a process called malting, which we describe later. Any pure, bipartite state is locally equivalent to a state in Schmidt form as
\be
\label{eq:init}
\ket{\psi^0} = \sum_n\alpha_n^0\ket{n}_{A_1}\ket{n}_{B_1},
\ee
where $\ket{n}_{A_1}$ ($\ket{n}_{B_1}$) denotes an $n$-photon Fock state in Alice's (Bob's) mode. This is the resource at the end of malting stage from which we will distill our final state using an iterative protocol. In the first step of the iteration, Alice and Bob combine two copies of the state $\ket{\psi^0}$ on two 50/50 beamsplitters. In the case that each party detects vacuum on one of the emerging modes from each beamsplitter, the resultant state in the other two modes is $\ket{\psi^1}$. Next, $\ket{\psi^1}$ is interfered with a fresh copy of $\ket{\psi^0}$ to produce $\ket{\psi^2}$ upon vacuum detection, and so on. At stage $i$ of the protocol, we combine $\ket{\psi^i}$ with $\ket{\psi^0}$ mode-wise on the beamsplitters, and detect vacuum, as in Fig.~\ref{fig:distillation_mashing}~(a), to produce the state
\be
\label{eq:gauss}
\!\!\ket{\psi^{i+1}} = \!\!_{A_1B_1}\!\bra{00}(U\!_{A_1A_2}\otimes U\!_{B_1B_2})\ket{\psi^i}_{A_1B_1}\otimes\ket{\psi^0}_{A_2B_2},
\ee
where $U_{ab}$ represents a 50/50 beamsplitter across modes $a$ and $b.$ If we denote
\be
\label{eq:expansion}
\ket{\psi^i} = \sum_n\alpha_n^i\ket{n}_{A_1}\ket{n}_{B_1},~~~ \mbox{for}~~ i=0,1,2...,
\ee
we obtain, from Eq.~(\ref{eq:gauss}), an iterative relation of the form
\be
\label{eq:map}
\alpha^{i+1} = \mathcal{M}\alpha^i,
\ee
where $\mathcal{M}$ depends on $\ket{\psi^0}.$ The convergence of this map is proven in~\cite{appendix}. Its behavior when different outcomes are detected in the mashing step, and its performance under decoherence is also discussed in~\cite{appendix}. A key innovation of our work is that we implement both mashing and malting steps in the same hardware. We call this hardware an entanglement distillery, and present a design for it which uses only four quantum memories.

\section*{An entanglement distillery}

Our distillery consists of four quantum memories to store entangled states during the protocol, and act as nonlinear and linear elements for generating and processing them. In fact, four photodetectors are the only other elements required to implement the distillery.

A quantum memory typically involves three modes, the input, the control, and a localised storage mode. The writing operation uses the control pulse to write the input into the storage mode, and the reading operation uses the control pulse to do the reverse. The storage mode $b$ is generally a matter degree of freedom, while the other two are optical. The simplest interaction for transferring a single excitation amongst three modes is of the form
\be
\label{eq:3wavemix}
\mathcal{H} \sim a^{\dag}b^{\dag}c + abc^{\dag}.
\ee
The beamsplitter required for the mashing step in Fig.~\ref{fig:distillation_mashing}~(b) can be readily achieved by setting the field $a$ to be classical. Note that one of the modes involved is optical, while the other is material. This allows us to exploit the best of both worlds: the optical modes for transferring information across the distillery, and the material modes for processing it.

Once copies of $\ket{\psi^0}$ are malted between memories $A_1$ and $B_1$, and between $A_2$ and $B_2,$ the matter modes in $A_2$ and $B_2$ are converted entirely into optical modes using strong control pulses. On Alice's side, the photons retrieved from $A_2$ are directed into $A_1$, and interfered with the matter mode via a 50/50 beamsplitter interaction, as in Fig.~\ref{fig:distillation_mashing}~(b). Correspondingly on Bob's side, the photons retrieved from $B_2$ are interfered with the matter mode in $B_1$. In the case that no photons are detected emerging from the ensembles, the state shared by Alice and Bob is projected into $\ket{\psi^1}$. The second iteration proceeds by malting another copy of $\ket{\psi^0}$ between $A_2$ and $B_2$, and interfering this with the matter modes in $A_1$ and $B_1$ as described above, which produces $\ket{\psi^2}$ provided both Alice and Bob detect vacuum again. Subsequent iterations are performed in the same way, by malting successive copies of $\ket{\psi^0}$ and mashing them into the state held in $A_1B_1.$  We note that this stage of the protocol requires a beamsplitter interaction with $T=0$ --- that is, $100\%$ retrieval efficiency from the memories $A_2$, $B_2$. The Raman scheme studied later makes this technically feasible \cite{appendix}.

Setting the mode $c$ in Eq.~(\ref{eq:3wavemix}) to be classical leads to a two-mode squeezing Hamiltonian, once again between an optical and material mode. This forms the first step of the malting process.

\section*{Malting}

Malting is a two-step process, depicted in Figs.~\ref{fig:distillation_malting}~(a) and (b). The first step is a two-mode squeezing interaction in memories $A$ and $B,$ held by Alice and Bob, denoted by the blue pulse in Fig.~\ref{fig:distillation_malting}~(a). These generate a pair of two-mode squeezed states of the form $\ket{\Phi} = \sqrt{1-\lambda^2}\sum_n \lambda^n\ket{n}\ket{n},$ where $\lambda$ is the squeezing parameter, and the matter mode of each memory is now entangled with its corresponding optical mode. The emitted photons are then directed over the channel connecting Alice and Bob, so that Alice receives Bob's photons, and Bob receives Alice's, as in Fig.~\ref{fig:distillation_malting}~(a). Each party now uses a control pulse (green pulse in Fig.~\ref{fig:distillation_malting}~(a)) to drive the same 50/50 beamsplitter interaction as used in the mashing step, so that Alice's photons are interfered with Bob's matter modes and vice versa. Finally, a photon counting detector placed behind each memory measures the optical mode emerging from the beamsplitter interaction. In the case that no photons are detected, the joint state of the two memories is again a two-mode squeezed state, now between the matter modes of Alice's and Bob's memories [Fig.~\ref{fig:distillation_malting}~(a)],
\be
\label{eq:tmss}
\ket{\Phi_{AB}} = \sqrt{1-\lambda^2}\sum_n \lambda^n\ket{n}_{A}\ket{n}_{B}.
\ee
In order to prepare a suitable non-Gaussian resource state $\ket{\psi^0}$, some non-Gaussian operation is now required. That is the aim of the second part of the malting process. We can perform any non-Gaussian operation on one or both of the modes of the two-mode squeezed state, and the memory provides a lot of versatility, but for the rest of this paper we will concentrate on photon subtraction, which has been studied in the context of entanglement distillation previously~\cite{Browne_ESP03,Eisert_BSP04,Opatrny_KW00,Cochrane_RM02,Olivares_PB03,Kitagawa_TSC06}.


Typically, photons are subtracted from optical modes using low-reflectivity beamsplitters and photon counters. This is a probabilistic process. In the same way, phonons can be subtracted from the matter modes by sending in weak control pulses and detecting the emission of a photon at the output (Fig.~\ref{fig:distillation_malting}~(b)). The advantage of using the matter modes to process information now becomes apparent, as the subtraction process can be tried repeatedly on the same copy of the initial state. By contrast, an optical implementation requires fresh preparation of the initial state if the subtraction fails. If after several weak control pulses (green pulses in Fig.~\ref{fig:distillation_malting}~(b)), a photon has been detected by both Alice and Bob, a successful subtraction on matter modes in both the memories has been heralded and our non-Gaussian resource state $\ket{\psi^0}$ is now ready. This completes the malting process. An example pulse sequence, as applied by Alice and Bob, is depicted in Fig.~\ref{fig:distillation_malting}~(b), where they require three and five attempts, respectively, to successfully implement subtraction. This is a fundamental advantage of a memory-based CV distillery, without any counterpart in free-space, finite-dimensional distillation schemes.

Each failed detection, however, alters the state. Since the initial state in the memory is a two-mode squeezed state, the quantum state after $f$ vacuum detections (over both arms) is still a two-mode squeezed state of the form of Eq.~(\ref{eq:tmss}), but with a squeezing parameter of $x=\lambda T^f,$ where $T$ is the effective transmissivity of the beamsplitter interaction. It is related to the reflectivity $R$ by $R^2 + T^2 =1.$ The larger the value of $T$ the better. Rather conveniently, $T$ can be made arbitrarily close to $1$ simply by reducing the energy of the subtracting control pulses. If we succeed in detecting photons at the photon counters in Fig.~\ref{fig:distillation_malting}~(b) after $f$ trials, our resource state takes the (unnormalized) form
\be
\label{eq:substate}
\ket{\psi^0} = \sum_{n=0}^{\infty}(n+1)\mu^n\ket{n}_{A_1}\ket{n}_{B_1},
\ee
with $\mu = xT^2 = \lambda T^{f+2}.$ Equating this with the entanglement of the initial two-mode squeezed state allows us to find the maximum number of tries $f_\mathrm{c}$ within which we must succeed if we are to have a net gain in entanglement. If $T = 1 -\eta,$ for small $\eta,$
\be
\label{eq:asymp}
f_\mathrm{c} \approx \left\lfloor\frac{\log(\lambda/R)}{\eta}\right\rfloor-2,
\ee
where $R$ is the real root of the equation $r^3 + (1-2\lambda)r^2 + (2-\lambda)r - \lambda =0$ ~\cite{appendix}. For typical parameters such as $\lambda=0.2$ and $T= 0.99,$ $f_\mathrm{c} = 60$. This means we have 60 attempts at subtracting excitations before there is no net advantage over the initial two-mode squeezed state in terms of entanglement. The dependence of $f_\mathrm{c}$ on $\lambda$ is weak, but a lower readout efficiency allows us more attempts at detecting photons.

\begin{figure}
\resizebox{8cm}{5.5cm}{\includegraphics{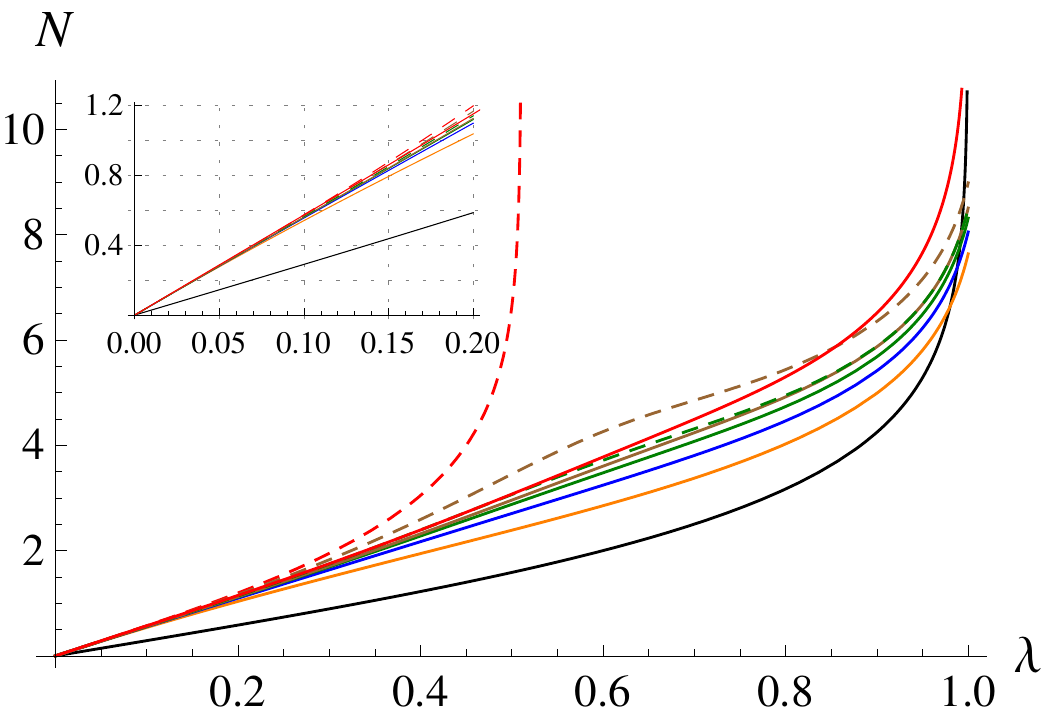}}
\caption{Entanglement, measured by the logarithmic negativity~\cite{p05}, of states after various steps of the distillation protocol. Black: Original two-mode squeezed state. Orange: Photon subtracted state in Eq. (\ref{eq:substate}). Solid lines: Our scheme. Dashed lines: Exponential scheme of \cite{Browne_ESP03}. Blue: 1 iteration, Green: 2 iterations, Brown: 3 iterations, Red: limiting state. Inset: Zoomed into the range $\lambda = [0,0.2].$ This illustrates the marginal difference in the yield of the two schemes, and the appeal of our exponentially improved scheme.}
\label{fig:compareschemes}
\end{figure}

\textit{Probability of successful malting:} The probability of success for the first step (Fig.~\ref{fig:distillation_mashing}~(b)), in which vacuum detections at each party herald the creation of $\ket{\Phi_{AB}}$, is $1-\lambda^2$, assuming perfect detectors. The probability of success of a memory-based subtraction at trial $f+1$ after $f$ vacuum detections is
\be
P_f = (1-T^2)^2x^2(1-\lambda^2)\frac{(1+\mu^2)}{(1-\mu^2)^3},
\ee
and the probability of succeeding by trial $f$ is $\overline{P}_f = \sum_{j=0}^f P_j.$  For the initial state $\ket{\psi^0},$ the results of the distillation scheme are presented in Fig. (\ref{fig:compareschemes}). As can be seen, the performance of our scheme is almost indistinguishable from that of the exponential protocol~\cite{Browne_ESP03}. Additionally, the state converges to the limiting state after a small number of iterations.

\begin{figure}
\resizebox{7.5cm}{5.5cm}{\includegraphics{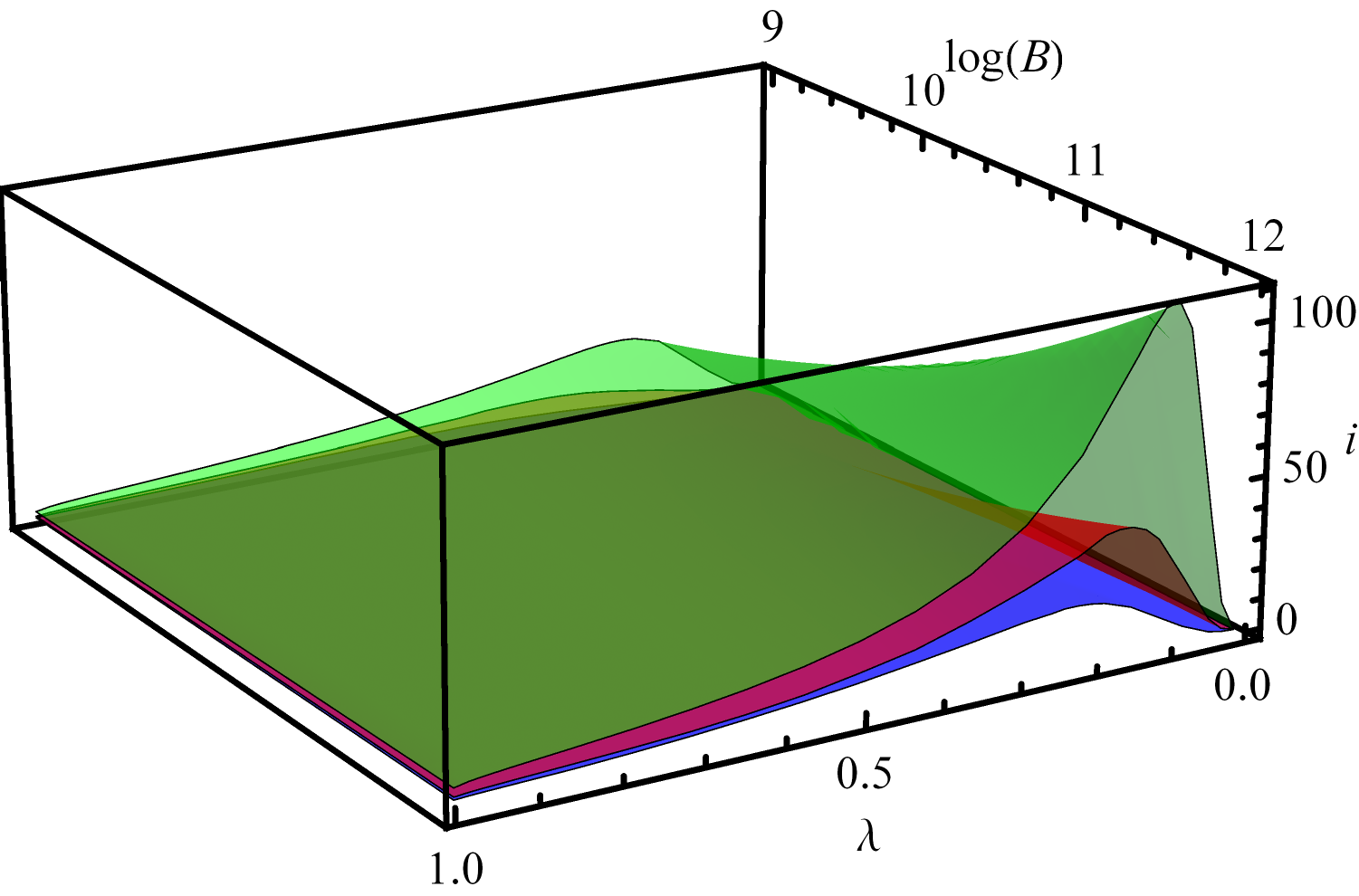}}
\caption{Number of iterations possible as a function of the time-bandwidth product $B$ and the initial squeezing $\lambda$, for various different values of $T$ used for photon subtraction: $T=0.8$ (Green), $T=0.9$ (Red) and $T=0.95$ (Blue). Smaller the value of $T,$ larger is the maximum number of possible iterations. This is in contrast to Eq.~(\ref{eq:asymp}), where the number of attempts decreases monotonically with $T$. This leads to a trade-off where one must choose the value of $T$ judiciously.}
\label{fig:pulses}
\end{figure}

In principle, our distillery could perform an arbitrary number of iterations, but in practice the number of iterations is limited by the finite storage lifetime $t_\mathrm{mem}$ of the memories. The useful lifetime of a memory is captured by its \emph{time-bandwidth product} $B=t_\mathrm{mem}/\tau$, which is the number of clock cycles, as defined by the duration $\tau$ of the control pulses, that fit within the lifetime of the memory. If $p^{s}_{\infty}$ is the probability of success of mashing two copies of the limiting state, the maximum number of iterations $i_m$ satisfies~\cite{appendix}
\be
\label{eq:tbw}
\frac{(i_m+1)f_\mathrm{c}}{\overline{P}_\mathrm{c}(p^{s}_{\infty})^{i_m}} \leq B.
\ee
For $\lambda = 0.15$, $T = 0.75$, and $B = 20000$, $i_m =54$. Fig.~(\ref{fig:pulses}) shows the number of iterations allowed for a broad range of parameters. For the instances we are interested in, a small number of iterations --- three, for example --- suffices to get close to the limiting case as shown in Fig.~(\ref{fig:compareschemes}). Note that a memory with a time-bandwidth product on the order of $B\sim10^3$ was recently implemented \cite{Reim:2010uq}, while $B\gtrsim 10^5$ is feasible with modest technical modifications, such as improved magnetic shielding. We next describe a quantum memory which can be used to implement a Hamiltonian of the form Eq. (\ref{eq:3wavemix}).

\section*{Raman memories}

\begin{figure}
\resizebox{7.5cm}{5.5cm}{\includegraphics{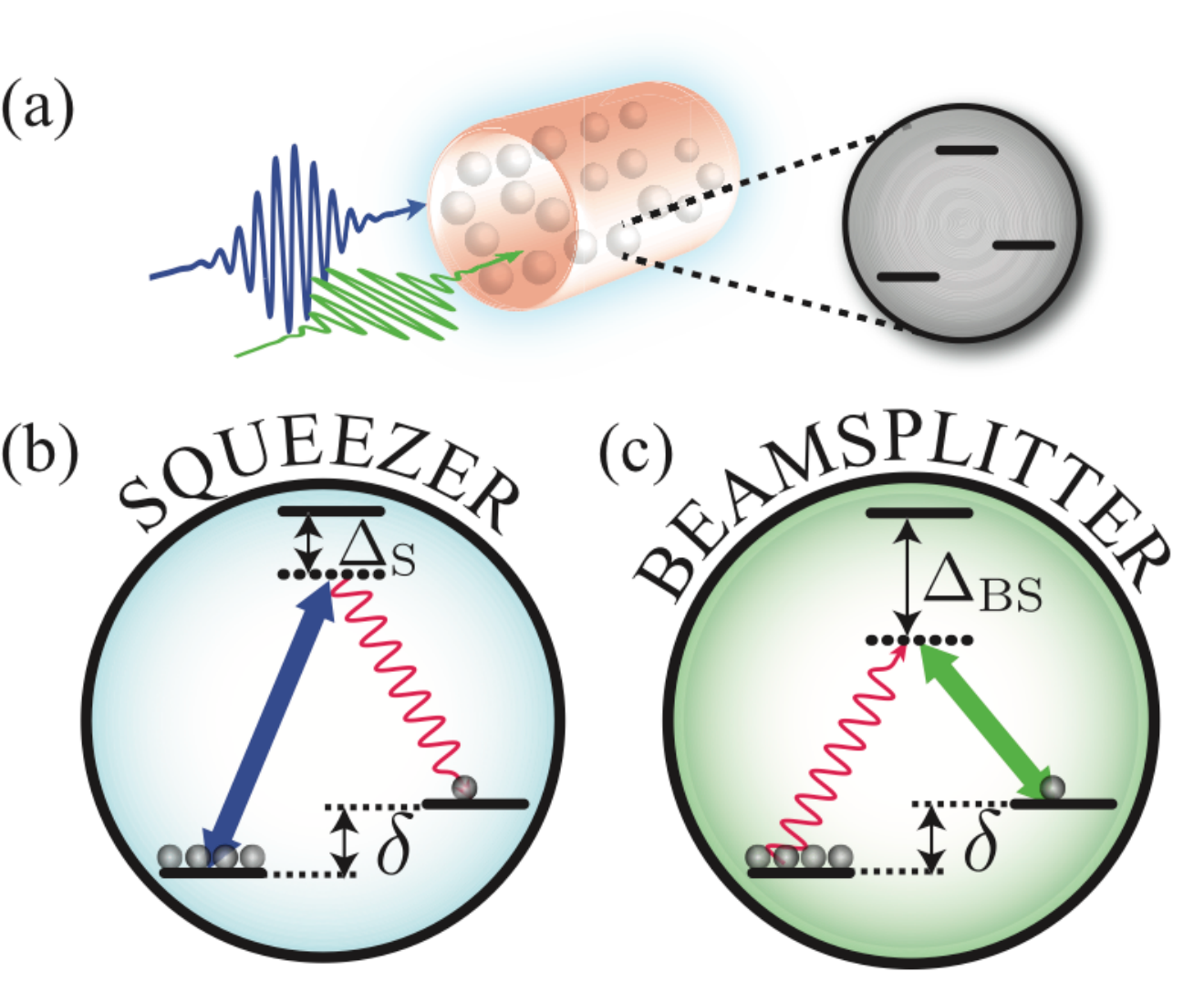}}
\caption{(a) An off-resonant Raman memory, comprised of an ensemble of atoms with $\Lambda$-type energy levels, provides a Hamiltonian of the form Eq.~(\ref{eq:3wavemix}).  (b) A two-mode squeezing Hamiltonian as in Eq.~(\ref{eq:hamS}) in which a strong, off-resonant \emph{pump} pulse with Rabi frequency $\Omega$, detuning $\Delta_\mathrm{S}$ from resonance and duration $\tau$ drives spontaneous Stokes scattering from the ground state. (c) Here, the pump pulse is replaced by an off-resonant \emph{control} pulse, with detuning $\Delta_\mathrm{BS}$, that couples the \emph{storage} state to the excited state. This results in an interaction as in Eq. (\ref{eq:hamBS}). Note that we have associated the same operator $a$ to both the anti-Stokes mode for the beamsplitter and the Stokes mode for the squeezer. In order to match the modes for the two types of interaction, we arrange the detunings such that $\Delta_\mathrm{BS} = \Delta_\mathrm{S} + \delta$, where $\delta$ is the Stokes splitting between the ground and storage states.
}
\label{fig:3levelsystem}
\end{figure}

Off-resonant Raman interactions in atomic ensembles provide a method for controllably coupling photons to non-propagating matter modes of atomic coherences known as \emph{spin-wave} excitations. The archetypal system consists of a vapour of three-level atoms with a $\Lambda$-type level configuration, in which a ground state is coupled to a long-lived storage state via an excited state. Two types of interactions can be implemented \cite{Hammerer:2010fk}, as shown in Fig.~(\ref{fig:3levelsystem}). The first is a two-mode squeezing interaction, in which the emission of a Stokes photon is accompanied by one of the atoms flipping from the ground to the storage state, producing a distributed excitation across the atomic ensemble, a spin wave. Stokes photons and spin-wave excitations are therefore produced in correlated pairs, and the interaction Hamiltonian has the form of a two-mode squeezer \cite{Wasilewski:2006th,raymer2}
\be
\label{eq:hamS}
\mathcal{H}_\mathrm{S} = C_\mathrm{S} a^{\dag}b^\dag +\mathrm{h.c.},
\ee
where $C_\mathrm{S}^2\approx \tau d\gamma \Omega^2/\Delta_\mathrm{S}^2 $ is the coupling strength, with $d$ the resonant optical depth of the ensemble and $\gamma$ the homogeneous linewidth of the excited state. Here $a$ and $b$ are annihilation operators for the Stokes and spin wave modes respectively.

The second kind of interaction has the form of a beamsplitter between optical and material modes. If there is a spin-wave excitation, it is converted into an anti-Stokes photon. On the other hand, an incident photon in the anti-Stokes mode will be absorbed and mapped into the spin wave mode. This allows for the storage and on-demand retrieval of optical pulses \cite{raman}. More generally, if photons and spin-wave excitations are simultaneously present, they will \emph{interfere}, precisely as two optical modes would at a beamsplitter; the energy in the control pulse determines the effective reflectivity of the beamsplitter interaction. The Hamiltonian for this interaction takes the form \cite{Nunn:2007wj}
\be
\label{eq:hamBS}
\mathcal{H}_\mathrm{BS} = C_\mathrm{BS} ab^\dag +\mathrm{h.c.},
\ee
where the coupling strength $C_\mathrm{BS}$ is the same as $C_\mathrm{S}$, except that the detuning $\Delta_\mathrm{S}$ is replaced with $\Delta_\mathrm{BS}$.  The beamsplitter ratio is $T \approx \sqrt{1 - C_{BS}^2}.$

\section*{Discussion}

The Herculean task of maintaining coherence in one part of a quantum device while another decoheres is the biggest roadblock to scalable quantum technologies. Quantum computation, communication, simulation, and metrology should all benefit immensely from the act of storing and processing quantum information is the same physical space. We have shown that protocols can be designed that cater to this requirement, and that systems exist to implement them. Quantum memories can be used not only to store, but process quantum information. They would also allow us to perform other local gaussian operations, such as squeezing~\cite{Yang_ZZsvLG11}, as well as non-gaussian operations. Our quantum distillery can produce high-quality entangled states between distant parties using just four quantum memories with imperfect storage efficiencies. We also showed that convergence to the limiting state can be achieved using realistic memories with limited storage times.

We have presented a protocol for CV entanglement distillation that is exponentially more efficient in spatial and temporal resources compared with previous schemes. The most attractive feature of our protocol is the ability to continue after failed probabilistic events. This is partly due to the use of continuous variables, which are more amenable to economic use than finite-dimensional systems like qubits. Critically, however, it is the localised nature of the continuous modes that really allows us to recycle the outcomes of failed attempts.

Preparation of entangled states is the starting point of most quantum information protocols.  Designing protocols which minimise the experimental resource requirements will be a key step in taking quantum information science into the realm of practical realisation.  Our work shows the potential for achieving this by making best creative use of existing components such as quantum memories.  We hope that this will encourage a stronger focus on the development of resource-efficient quantum protocols.

\section*{Acknowledgements}
We thank X.-M. Jin for several interesting discussions. This work was funded in part by EPSRC (Grant EP/H03031X/1), the European Commission (FP7 Integrated Project Q-ESSENCE, grant 248095, the EU-Mexico Cooperation project FONCICYT 94142), the US European Office of Aerospace Research and Development (Grant 093020), the Alexander von Humboldt Foundation, and the EU STREPs HIP and CORNER.



\section{appendix}

\subsection{Convergence of the mashing step of distillation protocol}

The mashing step, depicted in Fig.~\ref{fig:distillation_mashing}~(a), is written as
$$
\ket{\psi^{i+1}} = \!\!_{A_1B_1}\!\bra{00}(U\!_{A_1A_2}\otimes U\!_{B_1B_2})\ket{\psi^i}_{A_1B_1}\otimes\ket{\psi^0}_{A_2B_2}.
$$
Denoting
$$
\ket{\psi^i} = \sum_n\alpha_n^i\ket{n}_{A_1}\ket{n}_{B_1},~~~ \mbox{for}~~ i=0,1,2...,
$$
the mashing step can be expressed as
\be
\label{eq:iter}
\alpha_n^{i+1}=\frac{1}{2^n}\sum_{t=0}^n \left(\begin{array}{c}
                                          n \\
                                          t
                                        \end{array}\right)\alpha_{n-t}^i\alpha_t^0.
\ee
Each iteration, shown in Eq.~(\ref{eq:gauss}) and Fig.~\ref{fig:distillation_malting}~(a), maps the set of coefficients $\{\alpha_n^i\}_{n=0}^{\infty}$ into $\{\alpha_n^{i+1}\}_{n=0}^{\infty}.$ Hence, all the properties of the distillation scheme are encapsulated in the map (see Eq.~\ref{eq:map}) $\mathcal{M}_{jk}=\frac{1}{2^j}\left(\begin{array}{c}
                                          j \\
                                          k
                                        \end{array}\right)\alpha_{j-k}^0\Theta(j-k),$ where $\Theta(x)$ is the Heaviside step function. Expressed in matrix form
\be
\mathcal{M} = \left(
           \begin{array}{cccccc}
             \alpha_0^0 &  &  &  &  &  \\
             \frac{\alpha_1^0}{2} & \frac{\alpha_0^0}{2} &  &  \mbox{\Huge 0}&  &  \\
             \frac{\alpha_2^0}{4} & \frac{\alpha_1^0}{2} & \frac{\alpha_0^0}{4} &  &  &  \\
             \frac{\alpha_3^0}{8} & \frac{3\alpha_2^0}{8} & \frac{3\alpha_1^0}{8} & \frac{\alpha_0^0}{8} &  &  \\
             \vdots &  & \vdots &  & \ddots &  \\
              &  &  &  &  &  \\
           \end{array}
         \right).
\ee

It is easy to see that the fixed points of this map is given by $\alpha^0 = \{\lambda^n\}_{n=0}^{\infty}.$ Note that the iteration in Eq. (\ref{eq:iter}) leads to $\alpha^1_0=(\alpha^1_0)^2,$ whereby $\alpha^1_0 =1.$ Also, $\alpha^1_i$ is fixed, say to $\lambda$ all the remaining terms in the sequence are fixed. This proves that the fixed point is unique as well.

As we are ignoring normalization here, we can choose, $\alpha_0^0=1.$ Then the map has eigenvalues of the form $1/2^k,$ $k=0,1,2,\cdots,$ independent of the actual form of the $\alpha_n^0.$ Let us denote the (right) eigenvectors of $\mathcal{M}$ by $\ket{m_i}.$ Decomposing the initial state in terms of these eigenvectors as
\be
\ket{\psi^0} = \sum_k c_k \ket{m_k}.
\ee
After $i$ iterations, we get
\be
\mathcal{M}^i\ket{\psi^0} = \sum_k c_k \left(\frac{1}{2^k}\right)^i\ket{m_k}.
\ee
Note that the right eigenvectors do not form an orthonormal basis but still span the space. Thus,
\be
\lim_{i\rightarrow\infty}||\mathcal{M}^i\ket{\psi^0} - \ket{m_0}||=0.
\ee
This implies that the limiting state after an infinite number of iterations is the eigenvector corresponding to the eigenvalue 1.

Following Eq. (\ref{eq:expansion}), let us designate
\be
\lim_{i\rightarrow \infty} \alpha^i_n = \overline{\alpha}_n.
\ee
We note that $\alpha_0^{i+1} =\alpha_0^i\alpha_0^0,$ which implies
\be
\alpha_0^i = 1 \equiv \overline{\alpha}_0, ~~~\mbox{for}~~~i=0,1,2,\cdots.
\ee
From Eq. (\ref{eq:iter}), $\alpha_1^{i+1} = \frac{1}{2}(\alpha_1^i + \alpha_1^0).$ Iterating this recursion leads to
\be
\alpha_1^i = \alpha_1^0 \equiv \overline{\alpha}_1,~~~\mbox{for}~~~i=0,1,2,\cdots.
\ee
Also $\alpha_2^{i+1} = \frac{1}{4}(\alpha_2^i + 2(\alpha_1^0)^2 +\alpha_2^0),$ which leads to
\be
\alpha_2^i = \frac{2(\alpha_1^0)^2}{3}\left(1-\frac{1}{2^{2i+2}}\right) + \frac{\alpha_2^0}{3}\left(1+\frac{1}{2^{2i+1}}\right),
\ee
and so,
\be
\overline{\alpha}_2 \equiv \frac{2}{3}(\alpha_1^0)^2 + \frac{1}{3}\alpha_2^0.
\ee

Now we note a couple of features of this linear distillation scheme in general, after which we will specialize to specific input states. Our distillation protocol leaves the first two coefficients of the state unchanged, and only enhances the entanglement by modulating the higher Fock layers of the initial state. Also, in this linear scheme, the limiting state is determined by not just the $\alpha_1^0$ term of the initial state, but by all the terms in the initial sequence. This is the primary reason that makes the analytic derivation of the limiting state hard.

We now consider the a non-gaussian state as a resource, the candidate for our $\ket{\psi^0}$. The one we consider is a two-mode squeezed state with a photon subtracted from each arm. Such a state a given by Eq. (\ref{eq:substate}). For this particular input state,
\ben
\alpha_n^1 &=& \mu^n(n^2+3n+4)/4, \\
\alpha_n^2 &=& \mu^n(n^3+7n^2+24n+32)/32.
\een
It can be seen that the for this input state, $\overline{\alpha}_n = a_n\mu^n,$ with $0\leq a_n \leq 1.$ We can then rewrite the recursion relation in terms of the $a_n$ as
$$
a_n = \frac{1}{2^n}\sum_{t=0}^n \left(\begin{array}{c}
                                          n \\
                                          t
                                        \end{array}\right) (t+1)a_{n-t}.
$$

Its is interesting to note that outcomes different from that discussed above also lead to states with enhanced entanglement. Indeed, one could consider a mashing step as detecting $a_i$ and $b_i$ photons on Alice's and Bob's sides respectively leading to the iteration
$$
\ket{\psi^{i+1}} = \!\!_{A_1B_1}\!\bra{a_ib_i}(U\!_{A_1A_2}\otimes U\!_{B_1B_2})\ket{\psi^i}_{A_1B_1}\otimes\ket{\psi^0}_{A_2B_2}.
$$
The case we have considered in the simplest to analyze, where $a_i = b_i =0$ for all $i.$ More general cases, involving $a_i = b_i \neq 0, \forall i,$ as well as those involving different detection events in each iteration can also be analyzed, and adaptive schemes involving classical communication between the parties designed that lead to highly entangled distilled states. A more complete analysis of such distillation strategies is beyond the scope of this work, and will be presented elsewhere.

\subsection{Mashing in the presence of dephasing}

We discuss in brief the role of decoherence in our protocol. We envisage a scenario, where the distilled state at the end iteration $i$ has to wait in the memories while the source state $\ket{\psi^0}$ is malted. The state $\ket{\psi^i}$ will suffer dephasing in that time, and thereby loose its purity. To proceed further, we need to express the mashing step for mixed states. Let
\be
\rho^i = \sum_{kl;mn}\rho^i_{kl;mn}\ket{kl}\bra{mn}.
\ee
Then the mashing step corresponds to
\be
\rho^{i+1} = \!\!\bra{00}(U\!_{A_1A_2}\otimes U\!_{B_1B_2})\left(\rho^i_{A_1B_1}\otimes\rho^0_{A_2B_2}\right).
\ee
This can be expressed as an iteration on individual elements of the density matrix by
\begin{widetext}
\be
\rho^{i+1}_{PQ;RS}  = \frac{1}{2^{\frac{P+Q+R+S}{2}}}\sum_{p=0}^P \sum_{q=0}^Q \sum_{r=0}^R \sum_{s=0}^S \sqrt{\left(\begin{array}{c}
                                          P \\
                                          p
                                        \end{array}\right)\left(\begin{array}{c}
                                          Q \\
                                          q
                                        \end{array}\right)\left(\begin{array}{c}
                                          R \\
                                          r
                                        \end{array}\right)\left(\begin{array}{c}
                                          S \\
                                          s
                                        \end{array}\right)}
                                        \rho^i_{P-p,Q-q;R-r,S-s}\rho^0_{pq;rs}.
\ee

\begin{figure}
\resizebox{5.45cm}{4.5cm}{\includegraphics{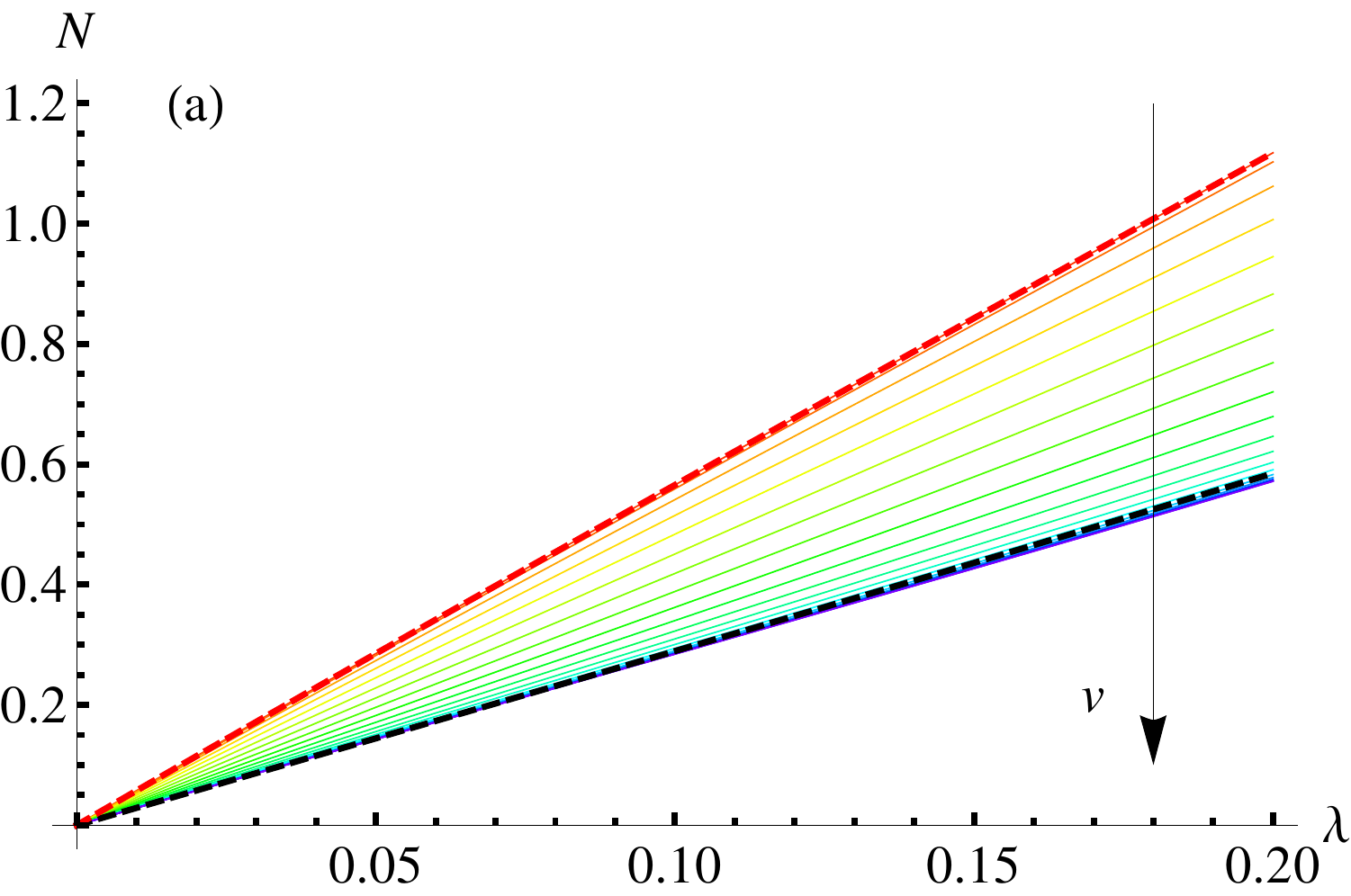}}
\resizebox{5.45cm}{4.5cm}{\includegraphics{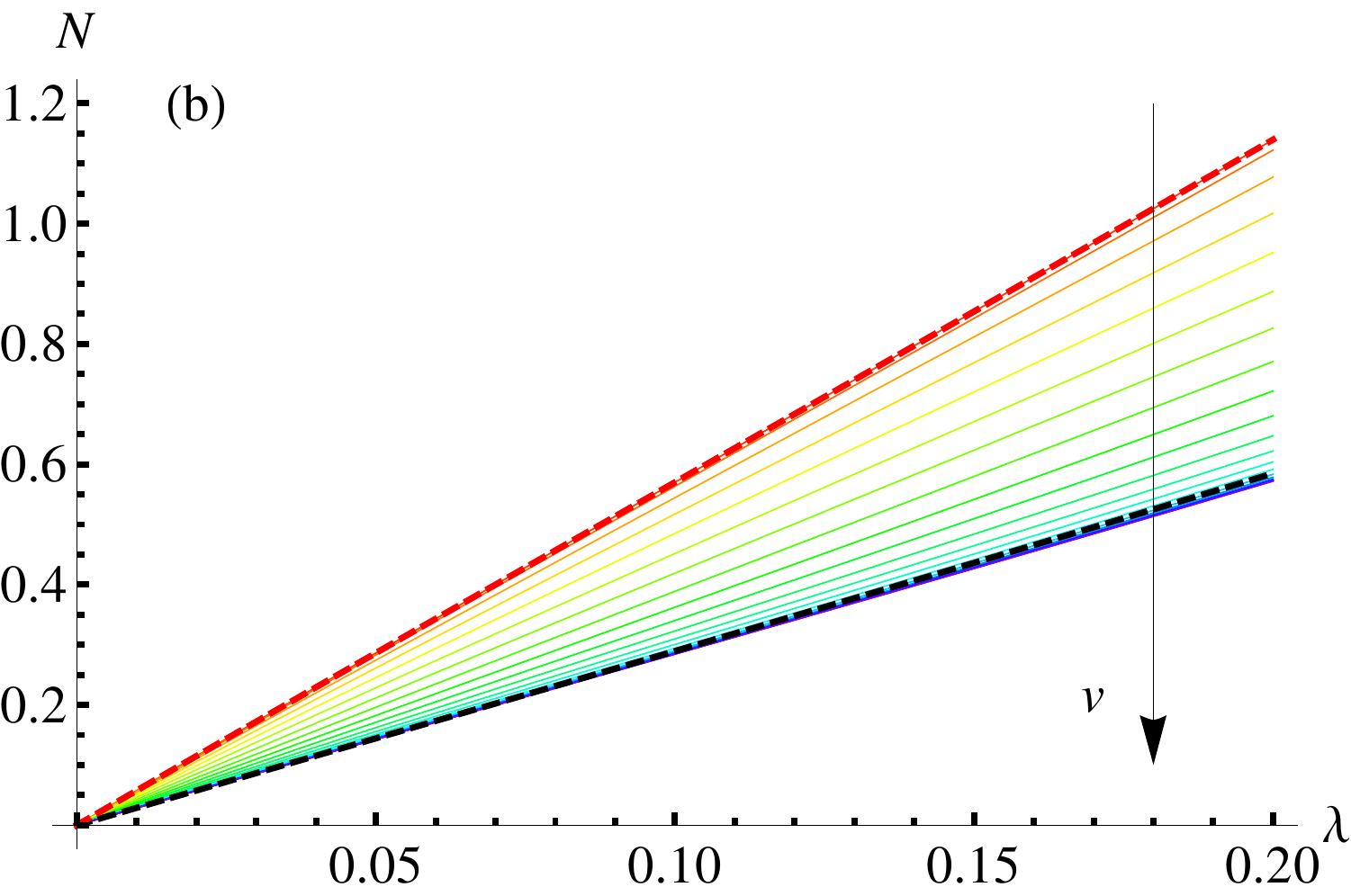}}
\resizebox{5.45cm}{4.5cm}{\includegraphics{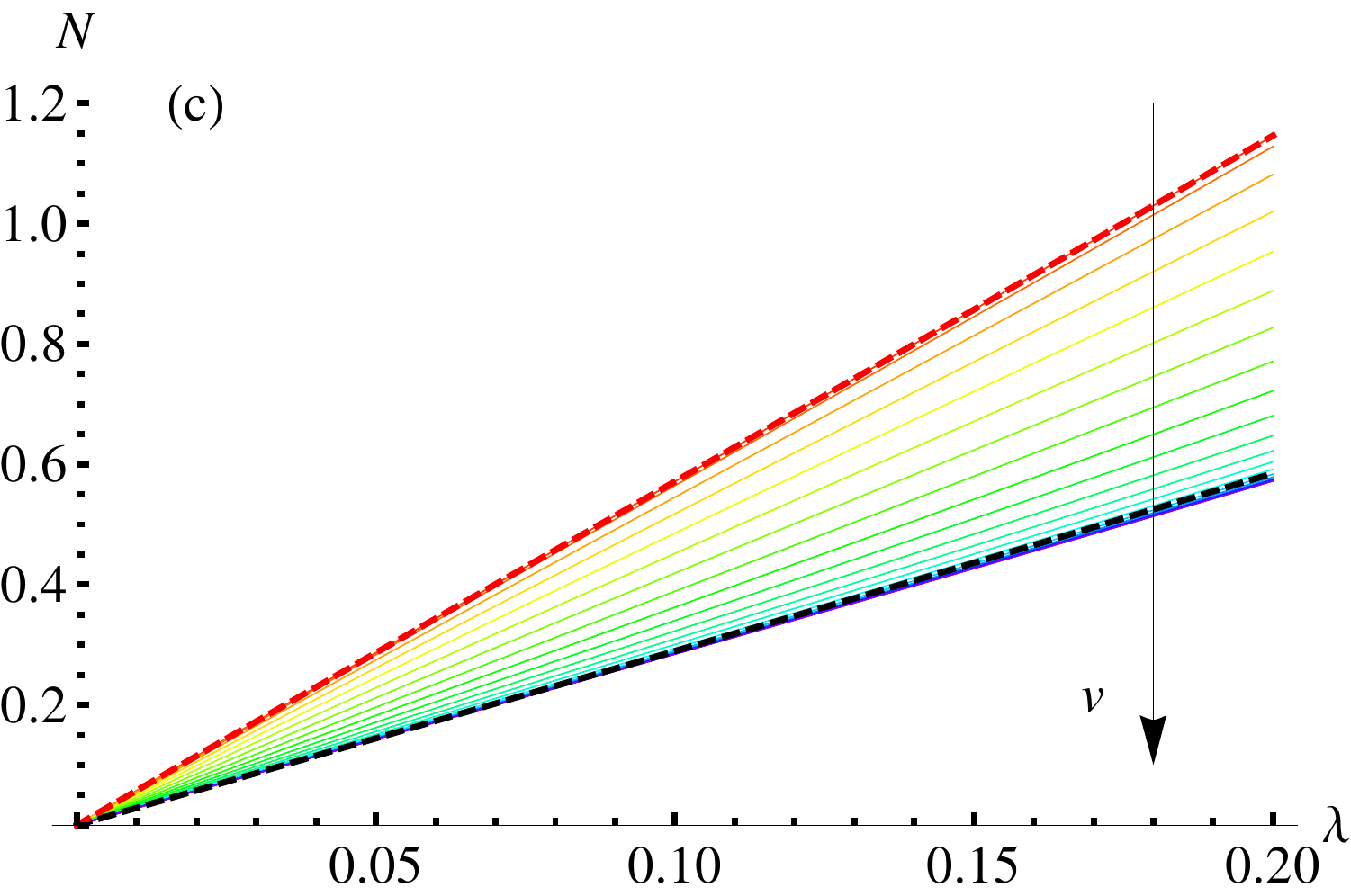}}
\caption{The logarithmic negativity as a function of the squeezing parameter for the iterations 1 to 3 from left to right. The thick, red, dashed lines denote the entanglement at the end of the iteration when there is not noise. The thick, blue, dashed lines denotes the entanglement of the initial two-mode squeezed state, and is identical for all 3 plots. The arrow pointing downwards indicates increasing values of $v$, from 0 to 2, with red denoting $v=0$, and going through the spectrum to violet denoting $v=2.$}
\label{fig:dephasing}
\end{figure}

\end{widetext}
We consider dephasing on the state $\rho$ phenomenologically as
\be
\bar{\rho}_{pq;rs} = \rho_{pq;rs}e^{-\frac{(p+q-r-s)^2}{2}v^2}.
\ee
$v$ denotes the amount of dephasing suffered by a state, and in the limit of $v \rightarrow \infty,$ the resulting state is completely diagonal, and consequently classical. In Fig.~\ref{fig:dephasing}, we show the effect of depashing on 3 iterations of our protocol. The largest value of $v$ that makes the process of distillation break even in the presence of noise is around $v = 2. $ Note that in the limit of complete dephasing of one of the states in an iteration, the state after the iteration still has as much entanglement as $\rho^0.$ This is another advantage of the linear protocol, which is much more susceptible to decoherence due to the exponential number of states involved. In a practical scenario, this would translate to requirements on the length of time one can wait between iterations, and other physical and material parameters.

\subsection{Maximum number of subtraction attempts}

The entanglement, as quantified by the logarithmic negativity~\cite{p05}, in the phonon-subtracted state in Eq.~(\ref{eq:substate}) is
\be
\mathcal{N}_f = \log_2\left[\frac{(1 + \mu)^3}{(1-\mu)(1+\mu^2)}  \right],
\ee
where $\mu = \lambda T^{f+2}.$ The logarithmic negativity of two-mode squeezed state with squeezing parameter $\lambda$ is
\be
\mathcal{N} = \log_2\left[\frac{1+\lambda}{1-\lambda}\right].
\ee
Equating the two above equations allows for a solution to $f=f_c$ as
\be
f_\mathrm{c} = \left\lfloor\frac{\log(R/\lambda)}{\log T}\right\rfloor -2 \approx \left\lfloor\frac{\log(\lambda/R)}{\eta}\right\rfloor-2,
\ee
where $R$ is the real root of the equation
\be
r^3 + (1-2\lambda)r^2 + (2-\lambda)r - \lambda =0.
\ee

\subsection{Probability of success of the malting step}

As is to be expected, $\overline{P}_f \geq \overline{P}_0, \forall f.$ Denoting the logarithmic negativity after $i$ iterations by $\mathcal{N}_i,$  a more reasonable figure of merit is the averaged entanglement gain given by $\overline{P}_\mathrm{c} \mathcal{N}_{f_\mathrm{c}}/\overline{P}_{0} \mathcal{N}_{0},$ and we have found this ratio to be larger than unity for a broad range of parameters, typically for $T>0.7$, independent of $\lambda$. As noted earlier, a larger value of $T$ allows for a larger value of $f_\mathrm{c}$. To get an idea of how a finite time-bandwidth product affects the number of iterations possible in our quantum distillery, we need to calculate the probability of success of an arbitrary iteration.

We begin by denoting the probability of success of the $i^\mathrm{th}$ iteration as $p^{s}_i.$ The combined probability of an entire sequence of $i$ successful iterations is then $\overline{P}_\mathrm{c}\prod_{j=1}^i p^{s}_j,$ where the $\overline{P}_\mathrm{c}$ is the probability of successfully preparing $\ket{\psi^0}$ within $f_\mathrm{c}$ trials as described above. Note that we are accounting for the worst-case scenario, whereby the initial state needs $f_\mathrm{c}$ attempts to be realized. As already mentioned, our memory-based subtraction scheme increases the probability of photon subtraction. Also, $i+1$ copies of this state are required for $i$ iterations of the protocol. Then the total number of memory operations is given by $(i+1)f_\mathrm{c}/\overline{P}_\mathrm{c}\prod_{j=1}^i p^{s}_j,$ which must be less than $B$.
\be
\label{eq:tbw}
\frac{(i_m+1)f_\mathrm{c}}{\overline{P}_\mathrm{c}\prod_{j=1}^{i_m} p^{s}_j} \leq \frac{(i_m+1)f_\mathrm{c}}{\overline{P}_\mathrm{c}(p^{s}_{\infty})^{i_m}} \leq B.
\ee
where for the first inequality we have used $p^s_i \geq p^s_{\infty},$ the probability of the step defined in Eq.~(\ref{eq:gauss}) succeeding in the limit of an infinite number of iterations. Since the limiting state is, by definition, invariant under Eq.~(\ref{eq:gauss}),
\be
p^s_{\infty} = \frac{1}{||\ket{\psi^0}||}= \frac{(1-\mu^2)^3}{1+\mu^2}.
\ee
Inequality (\ref{eq:tbw}) can be solved numerically.


\subsection{Efficient readout}
After malting a resource state $\ket{\psi^0}$ between the ensembles $A_2$, $B_2$, this must be mashed into the current entangled state $\ket{\psi^i}$ by retrieving it and interfering it with $\ket{\psi^i}$ in the ensembles $A_1$, $B_1$. This therefore assumes that we are able to retrieve a state with 100\% efficiency from the ensembles $A_2$, $B_2$, which is to say that we can implement a beamsplitter interaction with $T=0$. Fortunately, the Raman interaction allows for near perfect retrieval using \emph{multi-pulse readout}. That is, a train of several control pulses is directed into the ensembles. Each pulse may only achieve partial readout, but the combined effect of all the pulses enables, asymptotically, the complete extraction of the stored excitation. The retrieved state is now distributed over several temporal modes, but it remains coherent. Interfering this state with $\ket{\psi^i}$ now requires a single $T=1/2$ interaction in $A_1$, $B_1$. Since the incident field is delocalized over several time bins, the appropriate control field to couple it to the stored spin wave should have the form of a pulse train \cite{Nunn:2007wj}, but this is precisely what was used to drive the retrieval from $A_2$, $B_2$. Therefore the same train of control pulses can be re-used to mediate the interference. In fact, since only $T=1/2$ is required for interference, perfect modematching is not needed, and some attenuation of the control intensity can also be accommodated. This establishes the technical feasibility of each stage of our distillation protocol. Perfect storage is never required, and where near-perfect retrieval is desirable, it can be implemented easily with a train of several pulses --- the ability to interfere the resulting temporally delocalized state is also retained with this scheme.

\end{document}